\documentclass{nature}
\usepackage{graphicx}
\usepackage{booktabs}
\usepackage{caption}
\usepackage{amsmath}
\usepackage{amssymb}
\usepackage{lineno}

\title{Solid-state laser refrigeration of nanodiamond quantum sensors}

\author{Anupum Pant$^1$, R. Greg Felsted$^2$, Alexander B. Bard$^2$, Xiaojing Xia$^3$, Siamak Dadras$^{4,5}$, Kamran Shayan$^{4,5}$, Danika R. Luntz-Martin$^{5,6}$, Donald Mannikko$^{2}$, Ilia M. Pavlovetc$^{7}$, Stefan Stoll$^{2}$, Masaru Kuno$^{7,8}$, A. Nick Vamivakas$^{4,5,6,9}$  \& Peter J. Pauzauskie$^{1,2,10}$}

\begin{document}

\maketitle

\begin{affiliations}
 \item Materials Science and Engineering  Department,  University of Washington, Seattle, WA, USA
 \item Department of Chemistry, University of Washington, Seattle, WA, USA
 \item Molecular Engineering and Science Institute, University of Washington, Seattle, WA, USA
 \item Institute of Optics, University of Rochester, Rochester, NY, USA
 \item Center for Coherence and Quantum Optics, University of Rochester, Rochester, NY, USA
 \item Department of Physics and Astronomy, University of Rochester, Rochester, NY, USA
 \item Department of Chemistry and Biochemistry, University of Notre Dame, Notre Dame, IN, USA
 \item Department of Physics, University of Notre Dame, Notre Dame, IN, USA
 \item Materials Science Program, University of Rochester, Rochester, NY, USA
 \item Physical \& Computational Sciences Directorate, Pacific Northwest National Laboratory, Richland, WA, USA 
 
\end{affiliations}

\begin{abstract}
 The negatively-charged nitrogen vacancy (NV$^-$) centre in diamond is a remarkable optical quantum sensor for a range of applications including, nanoscale thermometry\cite{acosta2010temperature}, magnetometry\cite{rondin2014magnetometry,pelliccione2016scanned,schmitt2017submillihertz}, single photon generation\cite{kurtsiefer2000stable,aharonovich2016}, quantum computing\cite{childress2013diamond}, and communication\cite{nemoto2016photonic}. However, to date the performance of these techniques using NV$^-$ centres has been limited by the thermally-induced spectral wandering of NV$^-$ centre photoluminescence due to detrimental photothermal heating\cite{rahman2016burning}. Here we demonstrate that solid-state laser refrigeration can be used to enable rapid (ms) optical temperature control of nitrogen vacancy doped nanodiamond (NV$^-$:ND) quantum sensors in both atmospheric and \textit{in vacuo} conditions. Nanodiamonds are attached to ceramic microcrystals including 10\% ytterbium doped yttrium lithium fluoride (Yb:LiYF$_4$) and sodium yttrium fluoride (Yb:NaYF$_4$) by van der Waals bonding.  The fluoride crystals were cooled through the efficient emission of upconverted infrared photons\cite{seletskiy2016laser} excited by a focused 1020 nm laser beam. Heat transfer to the ceramic microcrystals cooled the adjacent NV$^-$:NDs by 10 and 27 K at atmospheric pressure and $\sim$10$^{-3}$ Torr, respectively. 
 The temperature of the NV$^-$:NDs was measured using both Debye-Waller factor (DWF) thermometry\cite{plakhotnik2014all} and optically detected magnetic resonance (ODMR)\cite{doherty2014temperature}, which agree with the temperature of the laser cooled ceramic microcrystal. Stabilization of thermally-induced spectral wandering of the NV$^{-}$ zero-phonon-line (ZPL) is achieved by modulating the 1020 nm laser irradiance. The demonstrated cooling of NV$^-$:NDs using an optically cooled microcrystal opens up new possibilities for rapid feedback-controlled cooling of a wide range of nanoscale quantum materials.
\end{abstract}

Nanodiamonds with negatively charged nitrogen vacancy centres  (NV$^-$:NDs) are widely known for their remarkable optical properties and have generated interest as key materials for quantum information science,\cite{childress2013diamond} nanoscale thermometry,\cite{acosta2010temperature} magnetometry,\cite{rondin2014magnetometry} and many other applications.\cite{schirhagl2014nitrogen} NV$^-$ spin systems have long (ms) coherence times at room temperature and their spin-readouts are readily accessible by visible lasers and microwave (MW) radiation. Optical levitation of nanodiamonds offers the additional advantage of extreme isolation, which is conducive to achieving ultra-high quality factor mechanical oscillation and longer spin coherence times for potential applications such as detectors in advanced cosmological observatories that study areas such as dark matter,\cite{riedel2013direct} quantum gravity,\cite{albrecht2014testing} and gravitational waves.\cite{arvanitaki2013detecting} 

Diamonds doped with nitrogen are prone to photothermal heating due to the background absorption of point defects including substitutional nitrogen (P$_1$) centres and both neutral (NV$^0$) and negatively charged (NV$^-$) nitrogen vacancy centres. This photothermal heating can lead to unintended particle ejection from an optical trap, or cause the graphitization and even burning of the trapped particle.\cite{rahman2016burning} The absence of conductive and convective pathways for removing excess heat from the optically-trapped particle \textit{in vacuo} magnifies the effect of photothermal heating.  Recent theoretical models\cite{kern2017optical} suggest that the direct radiative cooling of microdiamonds trapped in vacuum may be possible, but direct laser refrigeration of diamond remains to be demonstrated experimentally.

The recent cooling of fluoride-based ceramics to sub-cryogenic temperatures suggests that the indirect laser refrigeration of diamonds may be possible.\cite{melgaard2016solid} The radiative cooling within such fluoride ceramics is based on using a low-entropy laser beam to excite crystal field levels within Yb$^{3+}$ ions, followed by efficient anti-Stokes emission of high-entropy infrared photons.\cite{seletskiy2011local} Both microcrystals and nanocrystals of Yb:LiYF$_4$ and Yb:NaYF$_4$ have previously been shown to cool in vacuum\cite{rahman2017laser} and also in physiological media.\cite{zhou2016laser, roder2015laser}  However, to date solid state laser refrigeration has not been demonstrated to modulate the temperature of nanoscale quantum sensors.

Here we demonstrate the laser refrigeration of NV$^-$:NDs based on anti-Stokes luminescence from an underlying micrometer-scale ceramic substrate.  The use of micrometer-scale ceramics enables steady-state temperatures to be reached rapidly within microseconds to milliseconds\cite{crane2018} which is comparable to the electron spin coherence time of the NV$^{-}$ centre.\cite{stanwix2010coherence} Two direct measurements of the internal temperature of the NV$^-$:NDs are made using non-contact optical thermometry. First, a Debye-Waller factor (DWF) thermometry approach\cite{plakhotnik2014all} is used which is based on the the temperature-dependent ratio of the emission from the NV$^{-}$ centre’s zero-phonon to that of phonon overtones. Second, we use optically detected magnetic resonance (ODMR)\cite{plakhotnik2014all} to confirm the internal temperatures within the NV$^-$:NDs and show cooling by $\sim$10 K. A two-band differential luminescence thermometry (TB-DLT) approach is used to measure the internal temperature of the Yb:LiYF$_4$ crystals (Figure S1).\cite{patterson2010measurement} 

\begin{figure}
\includegraphics[width = 16cm]{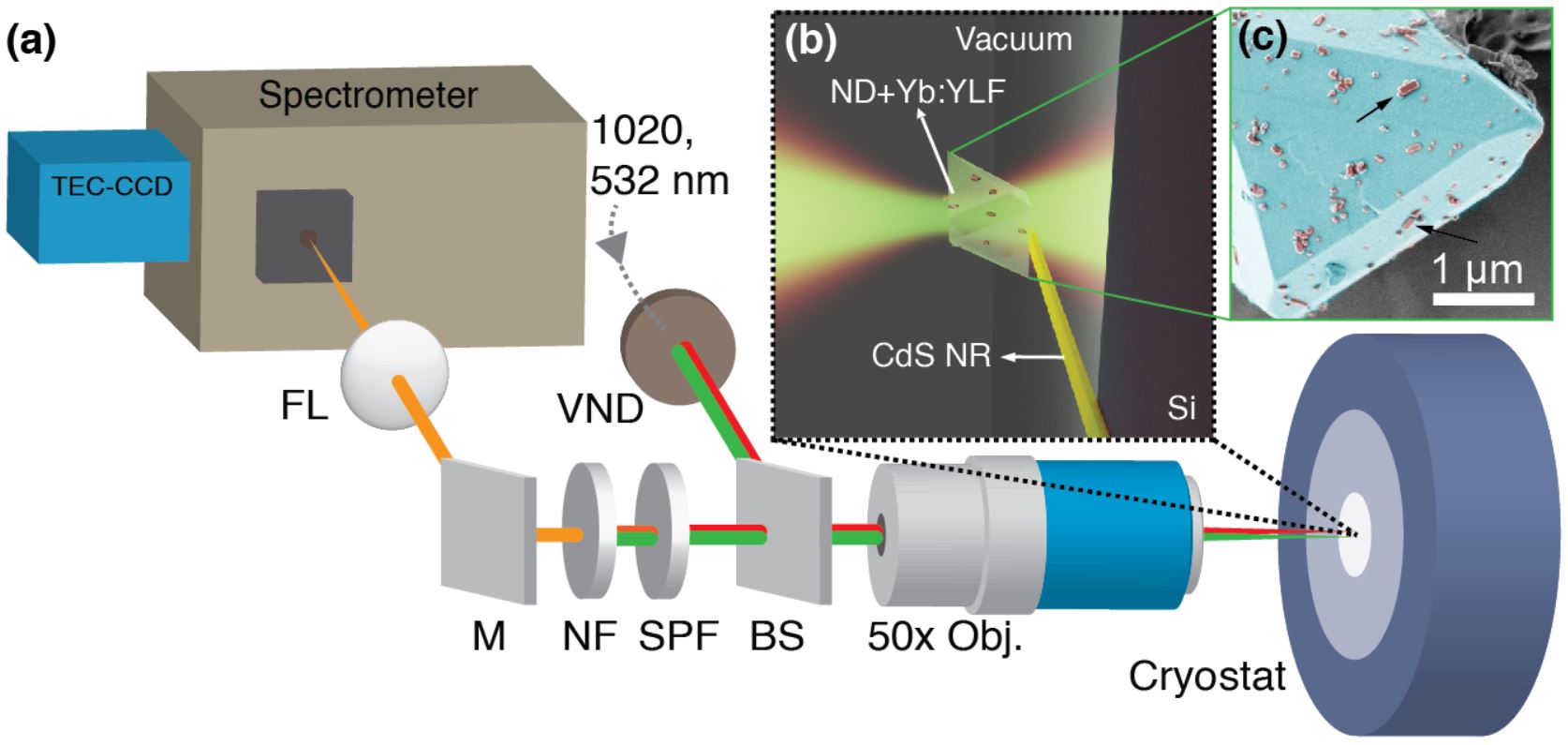}
\caption{(a) A schematic of the photoluminescence measurement setup. TEC-CCD, FL, VND, M, NF, SPF, and BS refer to thermoelectrically-cooled charge coupled device, focusing lens, variable neutral density filter, silver mirror, 532 nm notch filter, 1000 nm short pass filter, and beam splitter, respectively. (b) An illustration of the cadmium sulfide nanoribbon (CdS NR) cantilever device suspended in vacuum using a silicon substrate. A 10\% ytterbium doped lithium yttrium fluoride (Yb:LiYF$_4$) crystal with nanodiamonds (NV$^-$:NDs) dropcast on it is attached to the free end of the cantilever device. Subsequent inset on the right shows the (c) SEM of a representative Yb:LiYF$_4$ crystal with NV$^-$:NDs dropcast on it (colored for illustrative purposes). Representative NV$^-$:NDs on different facets of the the Yb:LiYF$_4$ crystal are indicated using black arrows.}
\label{setup}
\end{figure}

Thermal isolation of the laser cooled crystal was crucial for more efficient cooling for a given laser irradiance. Therefore, the cooling microcrystals with NV$^-$:NDs were placed at the end of a single-crystalline CdS nanoribbon cantilever that was suspended off of the edge of a silicon substrate. The small contact and cross-sectional area of the cantilever minimizes the heat conduction losses from the microcrystal. The CdS nanoribbon is $\sim$110 nm thick in the direction of light propagation, and is optically transparent at 1020 nm (and relatively transparent at 532 nm) which minimizes any additional heating due to its low absorption coefficient ($\alpha_\text{1020 nm}=6.7\times10^{-13}$ cm$^{-1}$, $\alpha_\text{532 nm}$=1362 cm$^{-1}$). An illustration of the device used for quantum temperature sensing is shown in Figure \ref{setup}b. A representative scanning electron micrograph of the Yb:LiYF$_4$ crystal coated with NV$^-$:NDs is shown in Figure \ref{setup}c. Nanodiamonds were coated onto the Yb:LiYF$_4$ crystal with a number density of $\sim$3.7$\pm$1 $\mu$m\textsuperscript{-2}.

\begin{figure}
\includegraphics[width = 16cm]{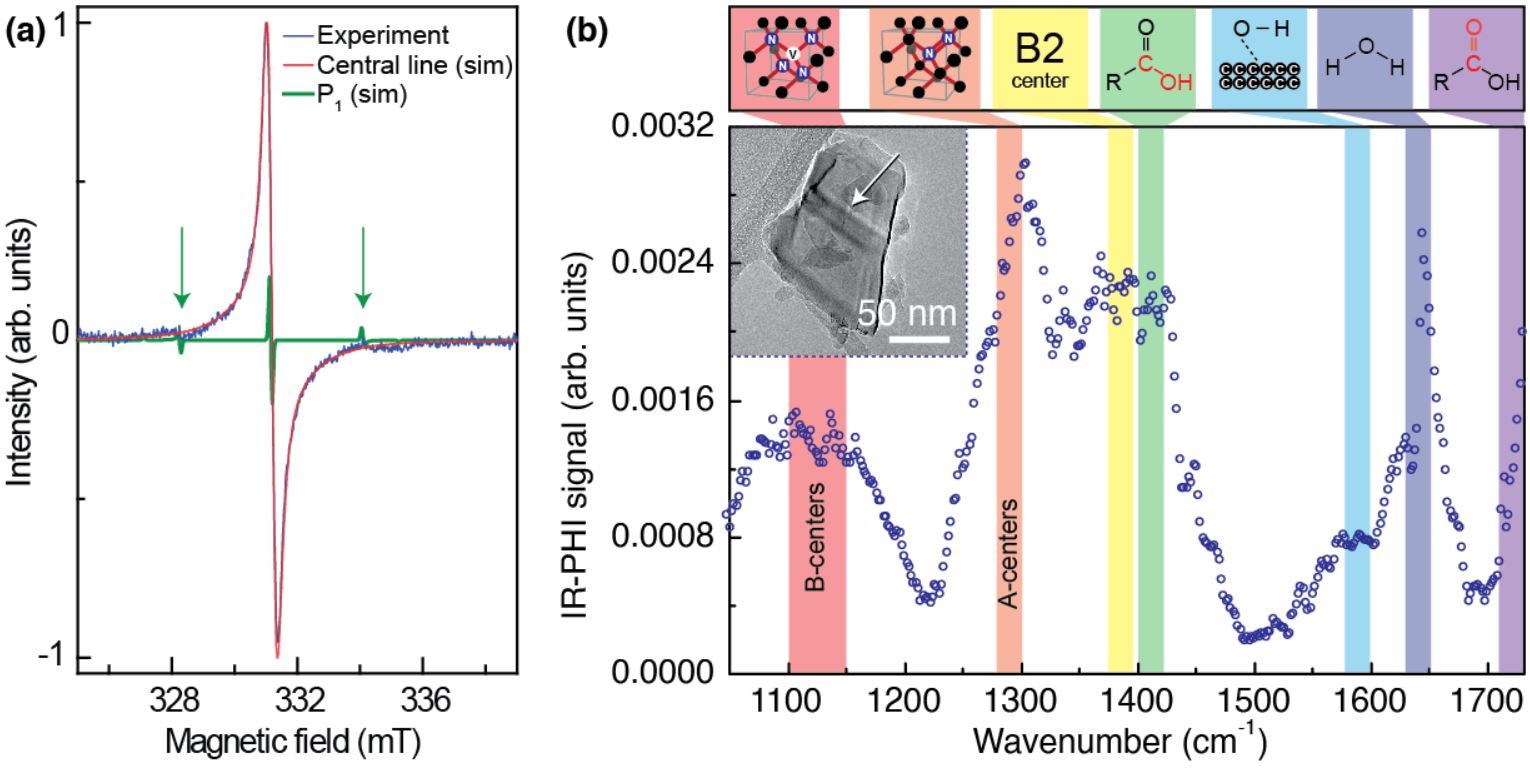}
\caption{(a) Room temperature EPR spectrum (blue) of the paramagnetic surface defects in an ensemble-NV$^-$:ND sample measured using a 9.30138 GHz microwave frequency of $\sim$1 mW power. The simulated central line and the simulated P$_1$ spectra are shown in red and green, respectively. (b) IR-PHI spectrum of a single nanodiamond sample (blue).
The peak assignments at wavenumbers (cm$^{-1}$) are 1100-1150: B-centres, 1280-1300: A-centres, 1380: B2 aggregates or C-H bond, 1415: carboxylic acid C-O-H stretch, 1590-1600: hydroxyl (O-H) on the surface, 1640: inclusions of water or water residue, 1730: carboxylic acid C=O stretch. Inset shows a TEM bright-field image of an individual NV$^-$:ND used in the shown IR-PHI measurements. }
\label{PL}
\end{figure}

In the NV$^-$ generation process, nitrogen initially enters the diamond lattice as a substitutional impurity during ion implantation (usually $\leq$200 ppm). Subsequent radiation damage induced by a variety of sources such as a beam of high-energy neutrons,
electrons, or ions is used to create vacancies.\cite{waldermann2007creating} Annealing of the irradiated diamond at T$\geq$850 $^{\circ}$C causes thermal diffusion of the vacancies. This results in the appearance of the desirable NV$^{-}$ complexes.\cite{waldermann2007creating} However, not all the nitrogen atoms are converted to the NV$^-$ centres, inducing a variety of undesirable defects. 

In addition to the photothermal heat generated from the broad phonon overtones of NV$^-$ centres, other surface or lattice defects within NV$^-$:NDs may contribute to photothermal heating.
To evaluate their presence, both electron paramagnetic resonance (EPR) and a superresolution infrared absorption spectroscopy technique called infrared photothermal heterodyne imaging (IR-PHI)\cite{li2017super} were used to make measurements on ensemble and single NV$^-$:ND particles, respectively. Figure \ref{PL}a shows the EPR spectrum of an ensemble of NV$^-$:ND. The strong central line consisting of the sum of two derivative Lorentzian line shapes is due to surface and other paramagnetic defects.\cite{osipov2017identification} The very weak outer lines (green arrows) are due to the substitutional nitrogen (P$_1$) centres in the NV$^-$:ND particles.\cite{cox199413c} The weakness of these lines is due to the low concentration of P$_1$ and the possible saturation of the signal.
The NV$^-$:NDs have $<$40 ppm of substitutional nitrogen (based on the manufacturer's specifications) which may also contribute to background photothermal heating.
IR-PHI spectra were collected from single NV$^-$:ND particles. A representative IR-PHI spectrum is shown in Figure 2b with the inset showing a transmission electron micrograph (TEM) image of an isolated NV$^-$:ND particle used in these measurements. 
The IR-PHI spectrum shows evidence of sizable concentrations of defects and also contains infrared transitions due to organic functional groups which passivate the NV$^-$:ND surfaces. In particular, the regions highlighted in red (1100-1150 cm$^{-1}$) and orange (1280-1300 cm$^{-1}$) in Figure 2b, show evidence for the presence of a significant amount of complex substitutional nitrogen point defects in the form of B-centres and A-centres, respectively. In addition, the presence of non-zero IR-PHI signal in the range of 1380 cm$^{-1}$ (yellow) may be due to the presence of B2-aggregates. These are complex objects in the \{100\} planes that can range from a few nanometers to micrometers in size and are formed by carbon and nitrogen interstitials.\cite{vasilyev2005interstitial} The presence of linear fringes across the centre of the NV$^-$:ND in the TEM image shown in the inset of Fig. 2b (white arrow) may also be the evidence of these defect centres. At this stage it is unclear which defect species beyond NV$^-$ centres contribute most to photothermal heating. 

\begin{figure}
\includegraphics[width = 16cm]{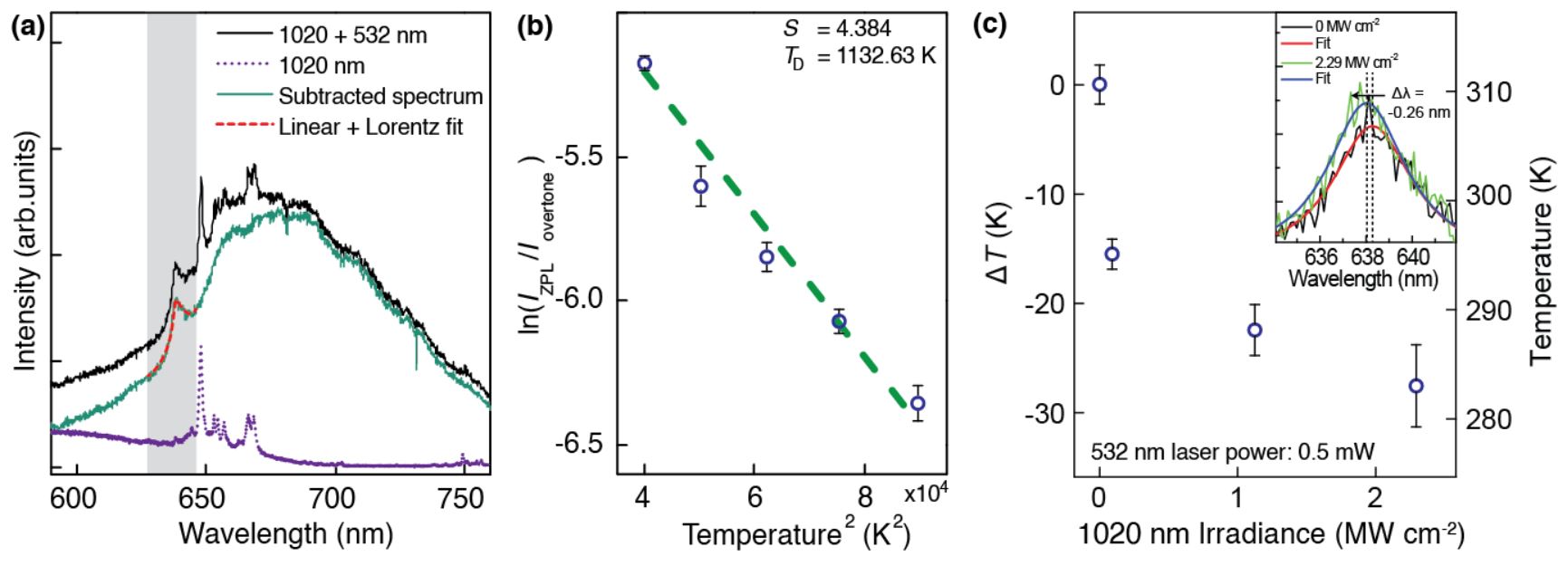}
\caption{(a) The composite PL spectrum, when both 1020 and 532 nm lasers are on, is shown in black. The contribution from the upconverted \textsuperscript{4}F\textsubscript{9/2} emission from Er$^{3+}$ impurity ions is shown using a dotted purple line. The subtracted spectrum and the fitting function are shown using a green and dashed red line, respectively. (b) The natural log of the DWF is plotted versus the temperature squared. A linear fit (dashed green line) to the data is used to extract $T_\text{D}$ and $S$. (c) The corresponding calibrated temperatures obtained from the measured value of ln($I_\text{ZPL}$/$I_\text{overtone}$) from the processed spectra are plotted for various laser irradiances of 1020 nm while the 532 nm laser is kept consistently at an irradiance of 0.073 MW cm$^{-2}$. The inset shows a close up of the background subtracted PL spectra in the ZPL region ($\sim$638 nm), and the Lorentzian component of the ZPL fit as the 1020 nm irradiance is increased from 0 and 2.29 MW cm$^{-2}$ are shown in red and blue, respectively. }
\label{DWF}
\end{figure}

Debye-Waller factor thermometry\cite{plakhotnik2014all} was used to measure the temperature of irradiated NV$^{-}$:ND quantum sensors by calibrating the ratio of the ZPL emission relative to the adjacent phonon side bands ($I_\text{ZPL}$/$I_\text{overtone}$). In Figure \ref{DWF}a, the black line shows the composite spectrum from the sample when both the 532 and 1020 nm beams were incident. Since the NV$^{-}$ emission spectrum was crucial in obtaining the temperature of the NV$^-$:NDs using the DWF technique, the background \textsuperscript{4}F\textsubscript{9/2} emission from the Er$^{3+}$ impurities (dotted purple line) overlapping with the emission of the NV$^-$ centre was subtracted from the composite spectrum (black line) at the respective intensity of 1020 nm laser used to obtain the emission profile from the NV$^{-}$ centres (green line). The ZPL was fit using a function consisting of a linear background and a Lorentzian around the ZPL (dashed red line) and the integrated intensity of the ZPL peak ($I_\text{ZPL}$), centred at 638 nm, was obtained from the amplitude of the Lorentzian component. The emission intensity from the phonon side bands ($I_\text{overtone}$) was obtained by subtracting the ZPL area from the integrated area of the spectrum. As the temperature of the cryostat was changed from 200-300 K, the DWF was calculated and plotted against temperature squared (Figure \ref{DWF}b). The data points were fit to the calibration function (dashed green line) discussed in Plakhotnik \textit{et al.}\cite{plakhotnik2014all} to obtain the Debye temperature ($T_\text{D}$=1132.6 K) and the electron-phonon coupling parameter ($S$=4.38). These values were used as a calibration for subsequent measurements. Figure \ref{DWF}c shows the calibrated temperature of the NV$^-$:NDs at various 1020 nm laser irradiances, where a continuous wave (CW) 532 nm laser at an irradiance of 0.073 MW cm$^{-2}$ was used as a probe beam to excite the NV$^-$ centres. As the 1020 nm laser irradiance was increased, an immediate drop in temperature and subsequent saturation trend was observed. A maximum temperature change of 27.7($\pm$3.8) K measured at 2.12 MW cm$^{-2}$ of 1020 nm laser irradiance using DWF thermometry agrees well with the temperature change of the Yb:LiYF$_4$ microcrystal (Figure S1). Therefore, the cooling power generated by Yb:LiYF$_4$ microcrystal in vacuum allowed net cooling of the attached NV$^-$:NDs by 27.7($\pm$3.8) K. The photothermal heating of NV$^-$:NDs from the 532 nm probe laser was measured to be 25.6 mK$\cdot\mu$W$^{-1}$ at zero 1020 nm irradiance. Heating from the probe laser can be reduced by using low probe laser irradiances on the order of 10 KW cm$^{-2}$ along with sensitive detectors. The inset of Figure \ref{DWF}c shows a close up of the background-subtracted PL spectra in the ZPL region ($\sim$638 nm). The ZPL fit at two 1020 nm irradiances is shown. As the irradiance is increased, the cooling induced blue-shift of the ZPL peak is observed from the respective fits. Tuning the 1020 nm irradiance allows for the stabilization of thermally-induced spectral wandering of the NV$^-$ centre's ZPL. A feedback loop can therefore be established so that the red-shifts in ZPL due to heating can be sensed optically and counteracted by increasing (cooling) pump laser intensities. The NV$^-$:ND temperature may be adjusted 3-6 orders of magnitude faster than a large cryostat's temperature considering the micro-millisecond timescales taken by the microcrystal to reach steady state temperatures.

\begin{figure}
\includegraphics[width = 16cm]{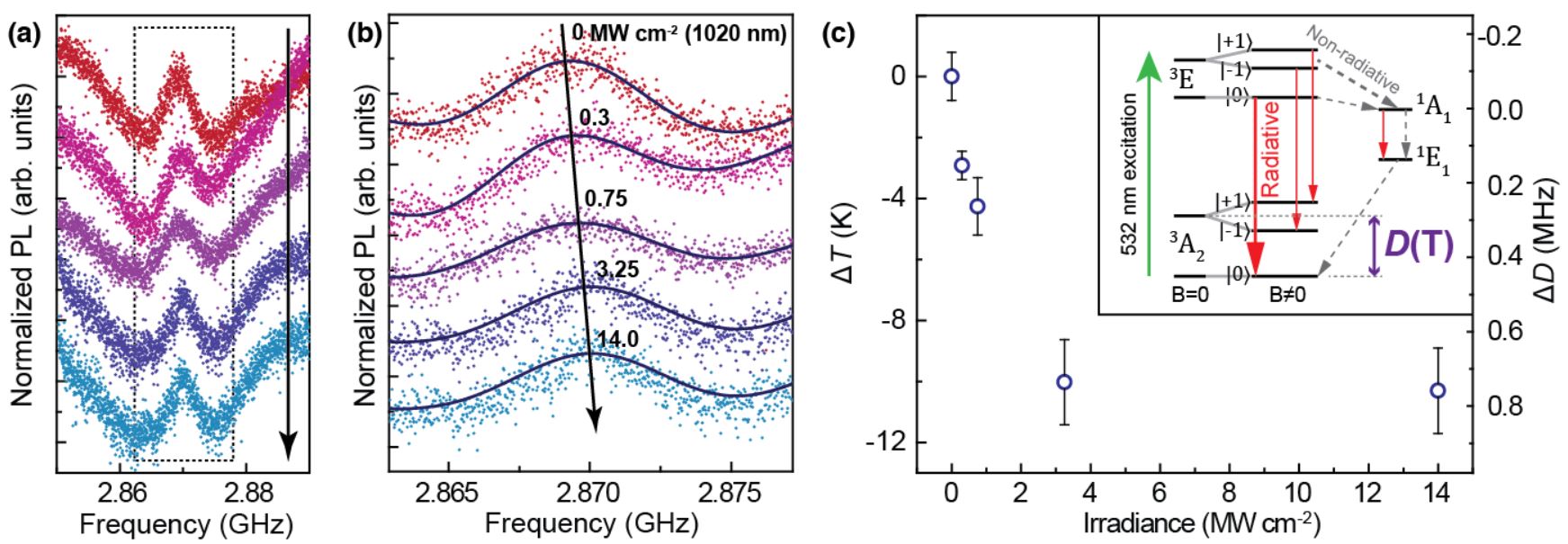}
\caption{(a) Full frequency range of ODMR measurements with two ($m_s$=0 $\leftrightarrow$ $m_s$=$\pm$1) resonance dips at $D \pm E$ (Eq. S2); at 1020 nm irradiance of 0, 0.3, 0.75, 3.25, and 14 MW cm$^{-2}$, increasing in the direction of the black arrow. The strained crystal lattice generates a pseudo-magnetic field that gives rise to a broken degeneracy and a splitting between the ground state spin sub-levels. (b) Variation of the zero-field splitting parameter $D$ as a function of the cooling laser irradiance. The ODMR signals, zoomed around 2.87 GHz, show a blue-shift in $D$ revealing the cooling of crystal with increased 1020 nm laser irradiance. The fits are shown in blue. The 1020 nm irradiance increases in the direction of the black arrow; (c) Blue-shift in the splitting parameter $D$ (right y-axis) and the corresponding reductions in the crystal's temperature (left y-axis) versus the irradiance of cooling laser. 
Inset shows the temperature dependent $D$-splitting parameter in the context of NV$^{-}$ centre's energy level system. The red arrows represent emission resulting from radiative decay. Heavier lines (as compared to a lighter lines) from the excited state to the ground state represents the stronger emission associated with $m_s$=0 transitions (versus the weaker emission  with $m_s$=$\pm$1). The weaker emission at $m_s$=$\pm$1 results from a stronger non-radiative coupling (heavier dashed grey line) to the singlet energy system through intersystem crossing.}
\label{ODMR}
\end{figure}

Alternatively, the cooling of NV$^-$:NDs can also be measured using ODMR\cite{acosta2010temperature} which does not require the subtraction of unwanted background emission from Er$^{3+}$ ions (Figure S2, S3). A hexagonal, oblate microcrystal of Yb:NaYF$_4$ grown via a modified hydrothermal synthesis\cite{zhou2016laser} was used to make the device shown in Figure S4 analogous to the device discussed above for DWF thermometry. For this device in ambient pressure, ODMR of the electron spins in NV$^-$ centres in nanodiamonds was used to measure the temperature of the NV$^-$:NDs. The readout of NV$^-$ spin systems is readily accessible by visible lasers and microwave (MW) radiation. Figure \ref{ODMR}a shows the ODMR spectra for low to high intensities (in the direction of the arrow) of the 1020 nm cooling laser. Figure \ref{ODMR}b shows the ODMR spectra zoomed around the centre frequency and the respective fits are shown using blue lines. As the laser irradiance is increased, a blue-shift in the zero-field splitting parameter ($D$) is observed in these spectra, indicating a reduction in the internal temperature of the crystal.\cite{acosta2010temperature} The net shifts of calibrated temperature measured (using the net shift in the splitting parameter and Eq. S2) at these intensities are shown in Figure \ref{ODMR}c. The corresponding increase in the splitting parameter is also shown in the right y-axis of Figure 4c. The splitting parameter increases as the laser irradiance increases from 0 to 3.25 MWcm$^{-2}$, corresponding to $\sim$10 K cooling in the crystal. A smaller temperature change compared to the measurements with Yb:LiYF$_4$ in vacuum are attributed to the different cooling crystal used (Yb:NaYF$_4$) and the presence of convective pathways due to ambient conditions. However, the experimental evidence shows that this cooling saturates by further increasing the laser irradiance, where the heating due to lattice vibrations counteracts the anti-Stokes cooling. The temperature change calculated based on the shifts in the ODMR signals are in close agreement with the ratio $dD/dT$=-74.2(7) kHz K$^{-1}$ from previous reports.\cite{acosta2010temperature} The ODMR-based temperature measurements are also consistent with recent reports\cite{fukami2019all} that fit the NV$^-$ centres' PL spectra around the ZPL. Fitting the ZPL spectra of the NV$^-$ centres with and without 3.25 MWcm$^{-2}$ cooling laser irradiation on the crystal also showed cooling (Figure S5). The calculated change in temperature agrees with the 10 K cooling found in the ODMR measurements (Figure S5).

In summary, we demonstrate the indirect laser refrigeration of nanodiamond quantum sensors via van der Waals attachment to ceramic laser-cooling microcrystals in vacuum and atmospheric pressure. We use complementary non-contact thermometry methods (both DWF and ODMR) to demonstrate solid state laser refrigeration as a rapid (ms) approach for controlling the temperature of nanoscale quantum materials near room temperature. A 27 K drop in the internal temperature of the NV$^-$:NDs was measured in vacuum. This allows for the stabilization of thermal spectral wandering of the NV$^{-}$ ZPL by tuning the 1020 nm irradiance. Laser refrigeration of NV$^-$:NDs will enable rapid feedback temperature control of quantum materials for a wide range of quantum sensing and communication applications. These results may also enable new experimental possibilities in quantum information science including single-beam laser trapping\cite{delic2020cooling,ashkin1986observation,pettit2017coherent}, precision sensing of temperature,\cite{acosta2010temperature} magnetic fields,\cite{rondin2014magnetometry} dark matter,\cite{riedel2013direct} and quantum gravity.\cite{arvanitaki2013detecting}

\begin{methods}

\subsection{Yb:LiYF$_4$ synthesis:}
The hydrothermal method used to synthesize single crystals of 10\% ytterbium doped lithium yttrium fluoride (Yb:LiYF$_4$) nanocrystals was performed following modifications to Roder \textit{et al.}\cite{roder2015laser} Yttrium chloride (YCl$_3$) hexahydrate and ytterbium chloride hexahydrate (YbCl$_3$) were of 99.999\% and 99.998\% purity, respectively. Lithium fluoride (LiF), lithium hydroxide monohydrate (LiOH$\cdot$H$_2$O), ammonium bifluoride (NH$_4$HF$_2$), and ethylenediaminetetraacetic acid (EDTA) were analytical grade and used directly in the synthesis without any purification. All chemicals were purchased from Sigma-Aldrich. For the synthesis of Yb:LiYF$_4$, 0.585 g (2 mmol) of EDTA and 0.168 g (4 mmol) LiOH$\cdot$H$_2$O were dissolved in 10 mL Millipore DI water and heated to approximately 80 $^\circ$C while stirring. After the EDTA was dissolved, 1.8 mL of 1.0 M YCl$_3$ and 0.2 mL of 1.0 M YbCl$_3$ were added and continually stirred for 1 hour. This mixture is denoted as solution A. Subsequently, 0.105 g (4 mmol) of LiF and 0.34 g (8 mmol) of NH$_4$HF$_2$ were separately dissolved in 5 mL Millipore DI water and heated to approximately 70 $^{\circ}$C while stirring for 1 h. This solution is denoted as solution B. After stirring, solution B was then added dropwise into solution A while stirring to form a homogeneous white suspension. After 30 minutes, the combined mixture was then
transferred to a 23 mL Teflon-lined autoclave (Parr 4747 Nickel Autoclave Teflon liner assembly) and heated to 180 $^{\circ}$C for 72 h in an oven (Thermo Scientific Heratherm General Protocol Oven, 65 L). After the autoclave cooled naturally to room temperature, the Yb:LiYF$_4$ particles were sonicated and centrifuged at 4000 rpm with ethanol and Millipore DI water three times. The final white powder was then dried at 60 $^\circ$C for 12 hours followed by calcination at 300 $^\circ$C for 2 hours in a Lindberg blue furnace inside a quartz tube.

\subsection{Yb:NaYF$_4$ synthesis:}
Yb:NaYF$_4$ crystals were grown using a hydrothermal process in a stainless-steel autoclave (Parr Instrument Company). YCl$_3$ (99.9\%), YbCl$_3$ (99.998\%), and ethylenediaminetetraacetic acid (EDTA, $>$99\%) were purchased from Sigma-Aldrich. NaOH was purchased from Fisher Scientific.  NaF (99.5\%) was purchased from EMD Chemicals.

\subsection{Sample fabrication:}
The edge of a clean silicon substrate was used to mount a CdS NR using a tungsten needle (World Precision Instruments) of a sufficiently small tip radius ($\sim$100 nm) under an optical microscope. The tungten tip was manipulated using a nanomanipulator (M\"{a}rzh\"{a}user-Wetzl\"{a}r). 
100 nm nanodiamonds doped with NV$^{-}$ centres (Adámas Nanotechnologies) were dropcast on the laser refrigeration microcrystals (Yb:LiYF$_4$ or Yb:NaYF$_4$). The particle density of NV$^-$:NDs on the surface of the crystals was 3.69 $\pm$ 1.00 particles/$\mu$m\textsuperscript{2}. Single crystals with NV$^-$:NDs on their surface were picked up and placed on the free-end of the cantilever using the same nanomanipulator.

\subsection{Infrared Photothermal Heterodyne Imaging (IR-PHI)}
IR (M-squared, OPO) and visible (532 nm, CNI Optoelectronics Tech.) lasers were used for IR-PHI measurements. Samples were diluted in DI water to achieve a desired concentration of 0.1 ng ml$^{-1}$ and were deposited on CaF$_2$ substrates via spin coating to ensure a final sample density below 0.1 $\mu$m$^{-2}$. The reflected 532 nm beam from the sample was then focused onto an auto-balanced detector, whose output was fed into a lock-in amplifier (Stanford Research Instruments) to measure the IR-PHI response at the IR laser repetition rate. IR-PHI spectra were obtained by measuring the IR-PHI response as a function of the IR OPO wavelength. A detailed description of the IR-PHI technique can be found in the reference.\cite{pavlovetc2020approaches}

\subsection{Electron paramagnetic resonance (EPR)}
The CW-EPR spectra were measured with a Bruker EMX spectrometer equipped with a Bruker SHQE resonator. The experimental sample was measured at room temperature with 1.0 mW microwave at 9.30138 GHz. The modulation frequency, amplitude, and sweep rate were were set to 100 kHz, 0.4 mT, and 0.358 mT s$^{-1}$, respectively.

\subsection{Debye-Waller Factor Thermometry (DWFT)}
The PL spectra were collected (Princeton instruments Acton SP500i and PROEM 512B CCD detector at -70 $^{\circ}$C). The area under the Lorentzian ZPL ($I_\text{ZPL}$) was obtained by fitting a composite (Linear + Lorentz) function to the spectrum in the small range of 625-645 nm (Figure \ref{DWF}a dashed red line). The area under the PSB ($I_\text{overtone}$) was obtained by numerically integrating the spectrum and subtracting the $I_\text{ZPL}$ from it. 

\subsection{Optically Detected Magnetic Resonance (ODMR)}
A frequency-doubled Nd:YAG laser (532 nm, 370 $\mu$W) was used for exciting the NV$^-$ centres by illuminating the crystal through a bottom objective (see Figure S3). Photoluminescence signal was collected using the same objective for PL mapping, ODMR, and spectroscopy. The sample was also illuminated from the bottom by a collimated while light beam and the back-scattered light was collected to locate and monitor the position of the sample. Microwaves (30 dBm) were applied on the sample using a loop antenna which delivered a sequence of 100 ms alternative on/off MW pulses (500 repetitions per location) to control the electron spin of NV$^-$ centres, and the frequency of the MWs was scanned over 2.85-2.89 GHz to perform ODMR measurements. 

\end{methods}

\bibliographystyle{naturemag}

\section*{Data availability}
The data that support the findings of this study are available from the corresponding authors on reasonable request.

\section*{Code availability}
code or algorithm used to generate results in this study are available from the corresponding authors on reasonable request.

\begin{addendum}

 \item A.P., X.X., I.P., M.K.,and P.P. gratefully acknowledge financial support from the MURI:MARBLe project under the auspices of the Air Force Office of Scientific Research (Award No. FA9550-16-1-0362). IP and MK additionally thank the NSF (CHE-1563528) for financial support. G.F., S.D., D.L.M, A.N.V. and P.P. acknowledge support from the ONR-BRC. Sample characterization was conducted at the University of Washington Molecular Analysis Facility, which is supported
in part by the National Science Foundation (Grant No. ECC-1542101), the University of Washington, the Molecular Engineering \& Sciences Institute, the Clean Energy Institute, and the National Institutes of Health.
 
 \item[Competing Interests] The authors declare that they have no competing interests.
 \item[Correspondence] Correspondence and requests for materials
should be addressed to Peter J. Pauzauskie. (email: peterpz@uw.edu).
\end{addendum}


\end{document}